# The Voltage Regulation of a Buck Converter Using a Neural Network Predictive Controller

Sepehr Saadatmand, Pourya Shamsi, and Mehdi Ferdowsi
Department of Electrical and Computer Engineering
Missouri University of Science and Technology
sszgz@mst.edu, shamsip@mst.edu, ferdowsi@mst.edu

*Abstract*—In this paper, a neural network predictive controller (NNPC) is proposed to control a buck converter. Conventional controllers such as proportional-integral (PI) or proportional-integral-derivative (PID) are designed based on the linearized small-signal model near the operating point. Therefore, the performance of the controller in the start-up, load change, or reference change is not optimal since the system model changes by changing the operating point. The neural network predictive controller optimally controls the buck converter by following the concept of the traditional model predictive controller. The advantage of the NNPC is that the neural network system identification decreases the inaccuracy of the system model with inaccurate parameters. A NNPC with a well-trained neural network can perform as an optimal controller for the buck converter. To compare the effectiveness of the traditional buck converter and the NNPC, the simulation results are provided.

## I. INTRODUCTION

Recently, the use of DC sources has increased rapidly in a vast area of application including renewable energy sources (RES) such as photovoltaic, electric vehicles, portable electronic devices such as cell phones and laptops, and aerospace. In some applications, the level of voltage needs to be changed to supply different loads [1]-[4]. Considering the enhancement in the fast-switching technology, power electronics converters are widely used in various applications. Therefore, the applications of the DC–DC converters have become more important [5], [6].

Semiconductor devices are the main core of power electronics converters, and they operate as electronic switches. The on–off mode causes nonlinearity in the system. The most common technique to control a DC–DC converter is based on conventional controllers such as the proportional-integral (PI) or the proportional-integral-derivative (PID). Conventional controllers are designed for the linear systems; hence, a linearized model in the neighborhood of the converter nominal operating point is used. Therefore, for a stable conventional controller, a significant change in the operating point might lead to system instability. Moreover, the other drawback of conventional controllers is that decreasing the overshoot percentage increases the rise time [7]-[10].

To enhance the transient response of power electronics converters, several studies have considered different control methods, such as fuzzy logic, model predictive control (MPC), neuro-fuzzy and sliding mode. A fuzzy logic control is presented in [11] for a DC–DC power converter in powered lighting system applications. The implementation of fuzzy-logic control algorithm for a DC–DC power converter using a microcontroller is explained in [12], [13]. The main advantage of the fuzzy logic is its behavior based on common policies and linguistics; hence, this method does not need the system model. Therefore, this method can perform well in the voltage regulation of DC–DC converters facing nonlinearity.

The fuzzy logic algorithm lacks formal analysis, and it is not considered a reliable controller by several authors [14]. Therefore, adaptive fuzzy control and model predictive control (MPC) have been studied as suitable replacements for the fuzzy logic technique [15]. The model predictive controller is a suitable controller for nonlinear systems, but its performance is highly dependent on the system model. Even if the system model is accurate, the uncertainties in the model parameters lead to inaccurate prediction. In other words, the model predictive controller overcomes the lack of analysis in fuzzy logic, but its dependency on the exact system model can extremely affect its performance.

Neural network–based controllers are powerful tolls when dealing with noise and uncertainties and are therefore widely implemented in applications such as supervised/unsupervised learning and reinforcement learning techniques. Several neuro-control techniques have been used in power electronics converters [16]-[22]. A neural network predictive controller (NNPC) is a suitable replacement for model predictive controller. This technique inherits both the advantages of the system model independency form fuzzy logic and the formal analysis of the model predictive controller. The system model of the implementation of a neural network predictive controller in a grid-connected synchronverter has been studied in [23].

The main contribution of this paper is to propose a neural network predictive controller for the voltage regulation of a step-down DC–DC converter. The rest of the paper is organized as follows. Section II discusses the mathematical model of the buck converter. The neural network predictive controller, the training process, and implementation are explained in Section III. The simulation results are provided in Section IV to evaluate the performance and the effectiveness of the proposed controller. Lastly, the conclusion is presented in Section V.

## II. BUCK CONVERTERS

Step-down converters are the power electronics converters that lower the level of voltage. The simplest form of a step-down converter is a buck converter. The output voltage of a buck converter is typically controlled by tuning the duty cycle of the pulse width modulation (PWM) signal. To avoid electromagnetic interference (EMI), the frequency of the PWM signal is typically fixed. The circuit diagram of a buck converter is shown in Figure 1, where *SW* and *D* are the power electronics switch and the diode, respectively. The input DC voltage is presented by $V_s$, the load resistance is shown by *R*, and the output voltage is $V_o$. A second-order low-pass LC filter is used to cancel out the switching frequency, where *C* is the filter capacitance and *L* is the filter inductance. The power electronics switch behaves as an electronic on–off switch; therefore, there are two different modes in the continuous conduction mode (CCM) operation: (i) when the switch is on, and (ii) when the switch is off, as shown in Figure 2. The first mode, when the switch is on, equation can be expressed as

$$\frac{d}{dt}\begin{bmatrix} i_L \\ v_c \end{bmatrix} = \begin{bmatrix} 0 & -L^{-1} \\ C^{-1} & -(CR)^{-1} \end{bmatrix}\begin{bmatrix} i_L \\ v_c \end{bmatrix} + \begin{bmatrix} L^{-1} \\ 0 \end{bmatrix}\cdot V_s \quad (1)$$

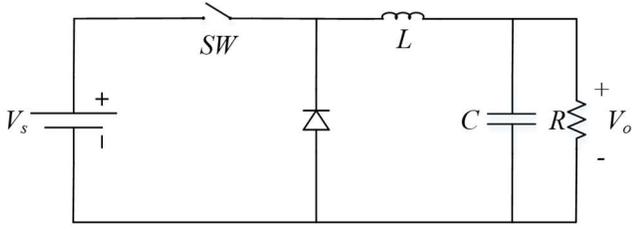

Figure 1. The circuit diagram of a buck converter

The state-space model of the system for the second mode when the switch is off, can be expressed as

$$\frac{d}{dt}\begin{bmatrix} i_L \\ v_c \end{bmatrix} = \begin{bmatrix} 0 & -L^{-1} \\ C^{-1} & -(CR)^{-1} \end{bmatrix}\begin{bmatrix} i_L \\ v_c \end{bmatrix} + \begin{bmatrix} 0 \\ 0 \end{bmatrix}\cdot V_s. \quad (2)$$

By applying the duty cyle and using the averaging model (1) and (2) can be written as

$$\frac{d}{dt}\begin{bmatrix} i_L \\ v_c \end{bmatrix} = \begin{bmatrix} 0 & -L^{-1} \\ C^{-1} & -(CR)^{-1} \end{bmatrix}\begin{bmatrix} i_L \\ v_c \end{bmatrix} + \begin{bmatrix} L^{-1} \\ 0 \end{bmatrix}\cdot d \cdot V_s. \quad (3)$$

The multiplication of the duty cycle and the input voltage illustrates the nonlinearity of the model. To linearized the model two approches have been made. The first approach is to find the equilibrioum and derive the linearized small-signal model. The second approach is to assume either the duty cycle or the input voltage are fixed. For example, the linearized state-space model under fixed voltage input can be written as

$$\dot{X} = AX + BU \quad (4)$$
$$Y = CX + DU \quad (5)$$

where *X*, *Y*, and *U* are the state vector, the input vector, and the output vector, respectively. Matrices *A*, *B*, *C*, and *D*

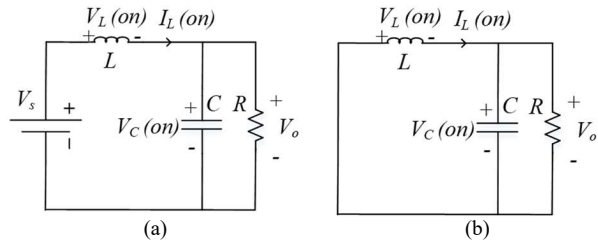

Figure 2. Circuit diagram of on-off mode

demonstrate the state-space matrices. The state-space parameters in buck converters can be defined as

$$X = [i_L \quad v_c]^T \quad (6)$$
$$A = \begin{bmatrix} 0 & -\dfrac{d}{L} \\ \dfrac{1}{C} & 0 \end{bmatrix} \quad (7)$$
$$B = [V_s \quad 0] \quad (8)$$
$$U = d \quad (9)$$
$$C = [0 \quad 1] \quad (10)$$
$$D = 0. \quad (11)$$

The drawback of this model is that the disturbance in the input voltage is not considered, and the drawback of the small-signal model is its dependency on the equilibrium operating point. If a nonlinear controller can be designed and implemented, the nonlinear form of the state-space model is more efficient and precise, which can be written as

$$\dot{X} = f(X, U, d) \quad (12)$$

where $f(\cdot)$ is a function of the state vector, control vector, and the duty cycle, which defines the derivative of the states with respect to the time as

$$f(X, U, \alpha) = AX + BdU. \quad (13)$$

The proposed state-space model defined in (13) can be used to design a nonlinear controller for the buck converter. However, to implement nonlinear controllers like the model predictive controller, this model needs to be accurate. Considering the parameter inaccuracy and uncertainties, Equation (13) fails to provide sufficient information for the optimizer block of the MPC.

### III. NEURAL NETWORK PREDICTIVE CONTROLLER

Optimization techniques have been used in a great variety of power electronics applications [23]-[26]. In this section, the neural network predictive controller is explained and analyzed. The model predictive control (MPC) technique optimally controls a system. The neural network predictive controller (NNPC) is a special case of the MPC that uses an artificial neural network to estimate the state-space function.

#### A. Model predictive control

The application of the MPC in power electronics started during the 1980s on the low switching frequency converters. However, the implementation of a MPC in high switching frequency is timely and expensive, and at that period, MPC methods were not very popular. After enhancement in producing a low-cost high-speed microcontroller, MPC schemes have garnered attention.

The MPC objective is to predict the behavior of a system under an optimal control policy in a specific time horizon. The prediction concept of the MPC can be explained by its state-space model in a discrete-time region as

$$x(k+1) = F(x(k), u(k)) \quad (14)$$
$$y(k) = G(x(k), u(k)) \quad (15)$$

where *x*, *y*, *u*, *k*, $F(\cdot)$, and $G(\cdot)$ are the discrete form of the state vector, output vector, control vector, time step, next state predictive function, and the output function, respectively. The

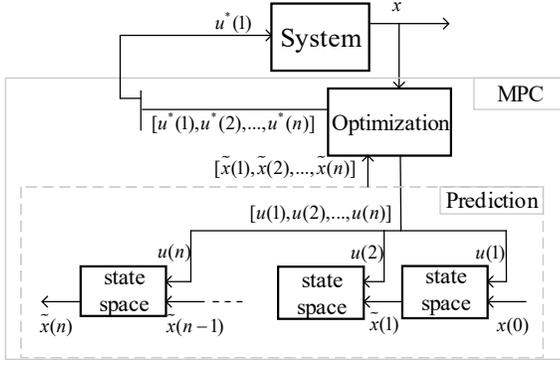

Figure 3. The block diagram of the model predictive controller

optimal control goal is to minimize a value or cost-to-go function $J$ defined as

$$J(x,u) = \sum_{k=1}^{N} \gamma^k \cdot C(x(k), u(k)) \quad (16)$$

where $C(\cdot)$ is a cost function, $\gamma \in [0, 1]$ is a discount factor to guarantee the divergence of the cost-to-go function, and $N$ is the maximum number of time step horizons. At each time step, the optimizer solves the optimization problem and provides a series of control vectors as the output. By applying the first vector of the optimal control series to the system, the next state will appear, and the process can then be repeated. The block diagram of a general MPC is shown in Figure 3. As depicted, the MPC includes two blocks: (i) the optimizer block, and (ii) the predictor block. The optimizer generates the control vector, and by feeding it to the predictor block using the state-space model, the series of the next states can be predicted and the cost-to-go function can be computed.

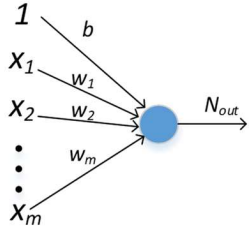

Figure 4. The block diagram of a single neuron with several inputs

The MPC for a system can be defined using the following steps:

1. modelling the system and implementing it as the predictor block
2. defining a cost function at each time step based on the state and the control at that time step
3. optimizing the discounted cumulative cost for that specific time horizon

The optimizer block can use several linear and nonlinear analytical and computational techniques to control the system to minimize the cost-to-go function.

Table I. Activation function

| | |
|---|---|
| Sigmoid | $f(x) = \sigma(x) = \dfrac{1}{1+e^{-x}}$ |
| TanH | $f(x) = \tanh(x) = \dfrac{e^x - e^{-x}}{e^x + e^{-x}}$ |
| Binary step | $f(x) = \begin{cases} 0 & \text{for } x < 0 \\ 1 & \text{for } x \geq 0 \end{cases}$ |
| Rectified Linear Unit (ReLU) | $f(x) = \begin{cases} 0 & \text{for } x \leq 0 \\ x & \text{for } x > 0 \end{cases}$ |

### B. Neural network structure

Ideally, when a system is simulated based on the exact model of that system, the MPC technique performs well. However, in real cases when the model of the system is approximated or the parameters are not accurate enough, having a robust predictive block is infeasible. To tackle these two drawbacks, an artificial neural network can be used as a system identifier to estimate the discrete form of the state-space model of the system. An artificial neural network is a network including one or multiple hidden layers with one or multiple neurons in each hidden layer, which mimics the behavior of a real neural network.

A single neuron structure is shown in Figure 4. The output of each neuron can be computed as

$$N_{out} = Act\left(b + \sum_{o=1}^{m} x_o \cdot w_o\right) \quad (17)$$

where $b$, $w_o$, $m$, and $Act(\cdot)$ are the neuron bias, the weight of the $i_{th}$ link, the number of neuron inputs, and the activation function, respectively. Based on the preferred output type, there are several activation functions. For example, a sigmoid activation function can be used to generate an output in [0, 1], or a tangent hyperbolic activation function might be used to have an output in [-1, 1]. Table I illustrates several activation functions.

By putting multiple neurons in a single layer, a single-layer neural network can be constructed. By cascading single-layer neural networks, a fully connected feedforward neural network can be formed, as shown in Figure 5. By using (18) to compute the neuron output, the general equation to compute the output of a multilayer neural network can be written as

$$N_{ij} = Act_j^i\left(b_j^i + \sum_{o=1}^{m_{i-1}} N_{(i-1)o} \cdot w_{oj}^i\right) \quad (18)$$

where $i$, $j$, and $N_{ij}$ are the number of layers, the number of neurons at that layer, and the output of the $j_{th}$ neuron at $i_{th}$ layer, respectively. The symbols $Act_j^i(\cdot)$ and $b_j^i$ are the activation function and the bias of the $j_{th}$ neuron at $i_{th}$ layer, respectively. The weight parameter between the $o_{th}$ input and $j_{th}$ neuron at $i_{th}$ layer is shown by $w_{oj}^i$. A feedforward neural network can be trained to map a set of input to a set of output. To train the neural network to map an input to an output, the weights and biases of the neural network need to be tuned. The gradient descent is the most common technique to update the network parameters and can be shown as

$$w_{oj}^i(k+1) = w_{oj}^i(k+1) - \alpha \frac{\partial P(k)}{\partial w_{oj}^i} \quad (19)$$

$$b_j^i(k+1) = b_j^i(k+1) - \alpha \frac{\partial P(k)}{\partial b_j^i} \quad (20)$$

where $\alpha$ is the learning rate, which determines the learning speed, and $P(\cdot)$ is the performance function of the network such as the cumulative square error.

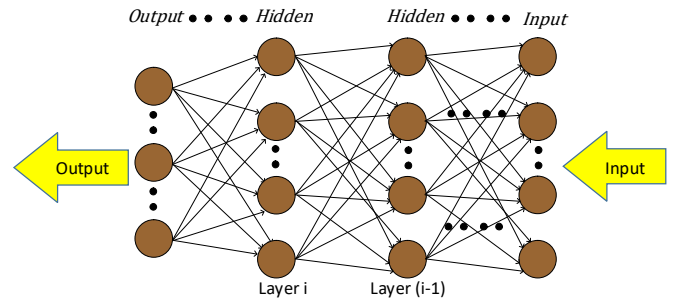

Figure 5. A fully connected multilayer feedforward neural network

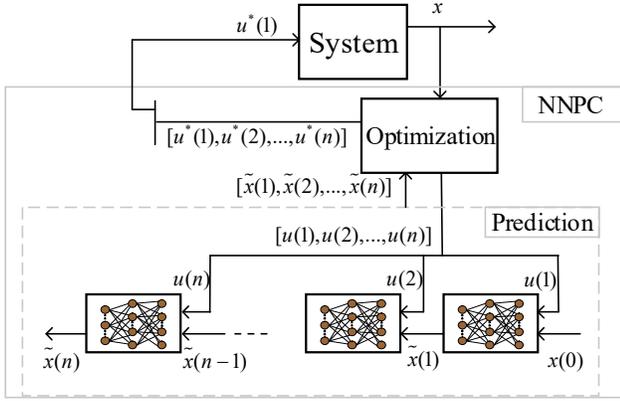

Figure 6. The block diagram of a neural network predictive controller

One of the most important applications of a fully connected feedforward neural network is the mapping and function estimation. With these applications, a neural network can be used to estimate the discrete state-space model of the system.

### C. Neural network predictive controller

As mentioned, the main drawback of the MPC is its dependency on the accuracy of the system model and the system parameters. A fully connected feedforward neural network can perform as the discrete state-space model of the system, which can map the current state and control to the state of the next step. The neural network needs to be trained prior to implementation. In other words, a set of training data needs to be prepared by running the system under random states and collecting the data at specific sample rates. After finishing the data acquisition process, the neural network can be trained to model the state-space of the system. The state-space model in the predictive block of the MPC can be replaced by the trained neural network, and the neural network predictive controller can be formed. The block diagram of a neural network predictive controller is shown in Figure 6. A single-layer neural network with seven neurons is selected to estimate the state-space model.

Table II. Buck converter parameters and information

| Parameter | Symbol | Value |
|---|---|---|
| Input voltage | $V_s$ | 48 V |
| Output voltage | $V_o$ | 12 V |
| Load resistance | $R$ | 6 Ω |
| Switching frequency | $f_{sw}$ | 75 kHz |
| Filter inductance | $L$ | 220 μH |
| Filter capacitance | $C$ | 10 μF |

## IV. SIMULATION RESULTS

To evaluate the proposed controller, an NNPC is implemented to control a buck converter. The block diagram of the proposed controller is illustrated in Figure 7. As shown in this figure, both PI and NNPC are implemented. The NNPC signal is disabled when the neural network is trained. In other words, the state signal goes to the PI controller, and the PI controller regulates the output voltage. The data is collected with the sample rate of 1 msec. After utilizing the buck converter with random references of output voltage and load current, the training data (including the state and the duty cycle at each time step,) is generated. The data set is a matrix (3×10000) that are collected in 10000 time steps. This matrix needs to be preprocessed to generate the input and output data. The input and output data to the neural network at each time step ($k$) can be expressed as

$$\text{Input}(k) = [\,i_L(k-1), v_c(k-1), d(k-1)] \quad (21)$$

$$\text{Output}(k) = [\,i_L(k), v_c(k)]. \quad (22)$$

The parameters of the buck converter is shown in Table II. The performance of the buck converter at start-up, load change, and the reference change of output voltage is evaluated and a comparison between NNPC and a PI controller is shown. The cost function to optimize is also defined as

$$C = \sqrt{(v_o - v_{ref})^2 + (i_L - i_{ref})^2} \quad (23)$$

### A. Start-up

Figure 8 illustrates the output voltage and the inductor current of the proposed buck converter during start-up. As expected, the system does not operate in its nominal operating point during transient time. Therefore, the performance of the PI controller includes voltage and current oscillations. However, the proposed NNPC optimally regulates the output voltage and the inductor current.

### B. Load change

To evaluate the performance of the proposed controller, a load change scenario from 6 Ω to 5 Ω is simulated. As previous simulations show, the PI controller does not function well when the performance of the buck converter is not near the nominal operating point. Figure 9 illustrates the output voltage and the inductor current for under both PI controller and NNPC for comparison.

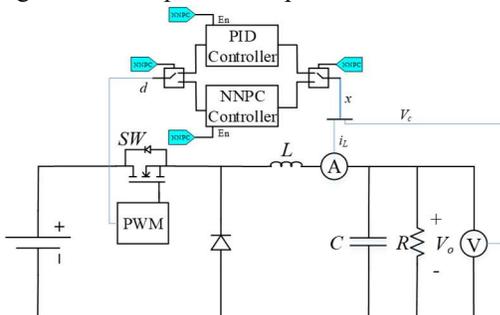

Figure 7. The block diagram of a NNPC-based buck converter

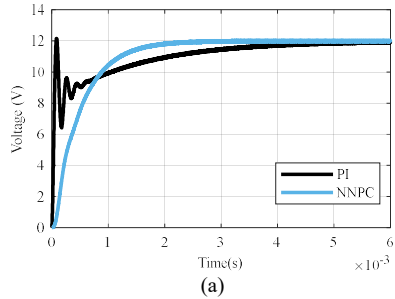
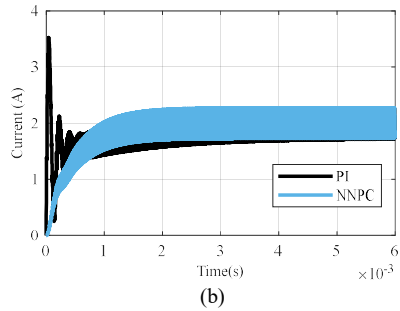

Figure 8. The performance of the buck converter in the start-up, (a) the output voltage, (b) the inductor current

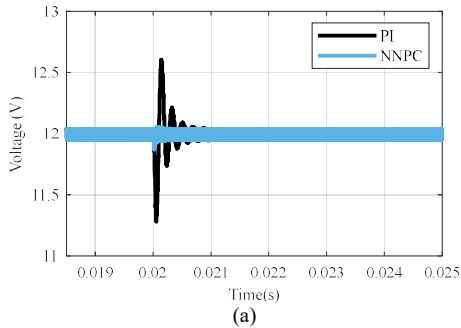
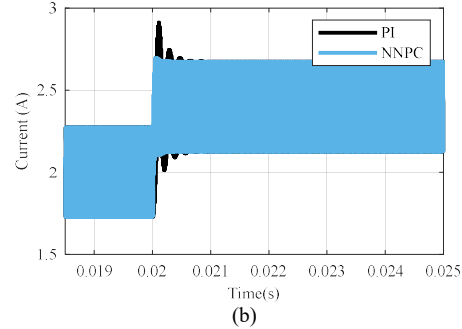

Figure 9. The performance of the buck converter at start-up, (a) the output voltage, (b) the inductor current

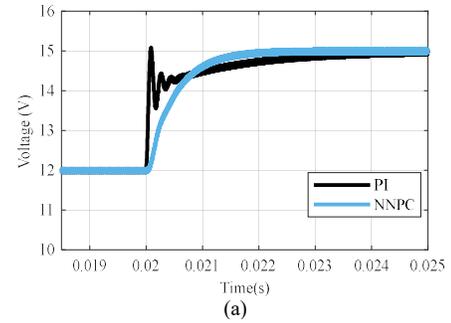
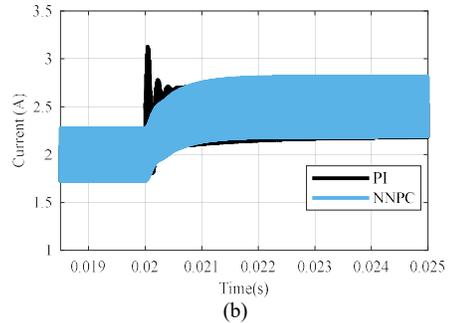
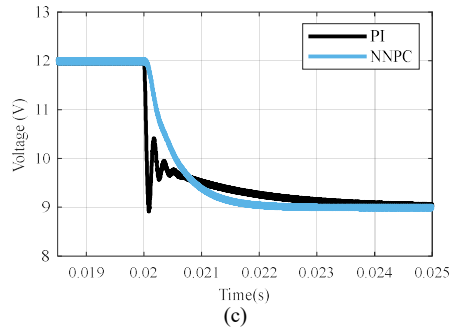
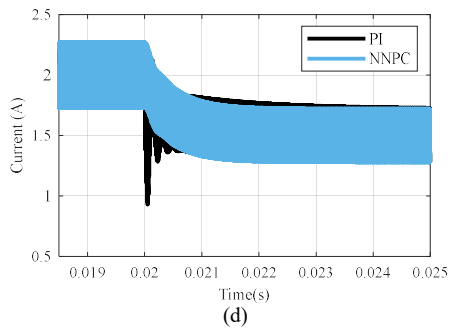

Figure 10. The performance of the buck converter in the reference voltage change, (a) the output voltage in reference voltage stepping down, (b) the inductor current in reference voltage stepping down, (c) the output voltage in reference voltage stepping up, (d) the inductor current in reference voltage stepping up

### C. Reference voltage change

To evaluate the performance of the proposed controller in reference voltage changes, two scenarios are considered. The first is when the reference voltage changes from 12 V to 15 V, and the second is from 12 V to 9 V. Changing the reference voltage alters the linearized state-space model based on which the PI controller is designed. Therefore, the performance of the PI controller is not optimal. However, the NNPC tracks the voltage reference with the minimum cumulative error at the optimal time horizon. Figure 10 depicts the voltage and the current output for both scenarios and compares the PI and the NNPC performance.

V. CONCLUSION

The penetration of DC–DC converters is rising rapidly due to the increase in penetration of renewable energy resources, electric vehicles, and portable electronic devices. Considering the enhancement in microcontroller technologies and the availability of cheap and fast microcontrollers, the model predictive controller attracted attention for its ability to overcome the drawbacks of the conventional controller. The MPC technique is highly sensitive to the model of the system, and inaccurate models or imprecise model parameters can severely affect the performance of the MPC. In this paper, a neural network predictive controller is proposed to control a buck converter. The proposed controller has the advantage of the MPC as a nonlinear controller, and the accurate estimation of the system model with the neural network overcomes the model dependency drawback of the MPC. As the simulation results show, the NNPC performs much better than the conventional controller since it is not based on a linearized model.